\documentstyle[twocolumn,aps,prl,epsf]{revtex}
\begin{document}
\title{
Tunneling Studies of Pseudogaps: a Comment}

\author{R.S. Markiewicz and C. Kusko$^*$} 

\address{Physics Department and Barnett Institute, 
Northeastern U.,
Boston MA 02115}
\maketitle

\pacs{PACS numbers~:~~71.27.+a, ~71.38.+i, ~74.20.Mn  }

\narrowtext

The recent observation of the pseudogap in tunneling measurements on 
Bi$_2$Sr$_2$CaCu$_2$O$_{8+\delta}$\cite{tu1,tu2,tu3} should prove of great value
in unravelling the mysteries of the `normal state' of the cuprates.  However, 
several issues in these papers require clarification.  Here, we discuss two 
important points.

\begin{figure}
\leavevmode
   \epsfxsize=0.33\textwidth\epsfbox{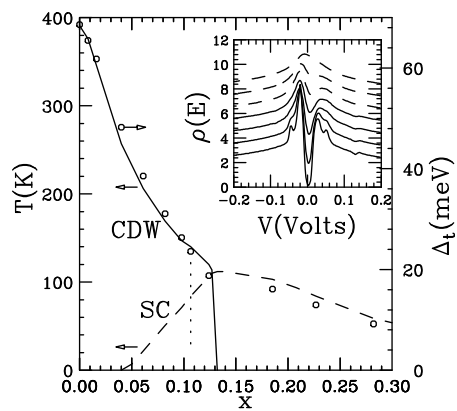}
\vskip0.5cm 
\caption{Phase diagram of pinned Balseiro-Falicov model. Circles = net tunneling
gap, $\Delta_t$.
Inset: Tunneling spectra of a density-wave superconductor, using 
parameters of dotted line in main frame.  Temperatures (from top to bottom) = 
130, 110, 90, 80, 70,
50, and 30K (dashed lines: T above the superconducting transition temperature).}
\label{fig:1}
\end{figure}

First, the gaps observed in the quasiparticle tunneling spectra are assumed to
be superconducting gaps, and taken as evidence that the pseudogap is caused by
superconducting fluctuations.  However, a normal-state gap (due, e.g., to 
charge or spin density waves) will
also show up in the tunneling spectra\cite{Gab}.  For illustrative purposes,
we use the pinned\cite{RM8a} Balseiro-Falicov (BF)\cite{BFal} model of 
competition between a charge density wave (CDW) and (s-wave) superconductivity 
(SC), which gives a good account of the doping dependence of the 
pseudogap\cite{RMPRL} and is a simple model for striped phases\cite{Pstr}.  For 
a pure CDW, the spectral function is of BCS form:
\begin{equation}
A(k,\omega )=2\pi [u_k^2\delta(\omega -E_{k+})+v_k^2\delta(\omega -E_{k-})],
\label{eq:1}
\end{equation}
with
$u_k^2=1-v_k^2=(1+\epsilon_{k-}/ \tilde E_k)/2$,
$E_{k\pm}=(\epsilon_{k+}\pm\tilde E_k)/2$,
$\epsilon_{k\pm}=\epsilon_k\pm\epsilon_{k+Q}$ and $\tilde E_k=\sqrt
{\epsilon_{k-}^2+4G_k^2}$, where the nesting vector $Q=(\pi ,\pi )$, and
the gap $G_k$ and dispersion $\epsilon_k$ are defined in Refs. \cite{RM8a,BFal}.
Figure~\ref{fig:1} shows the calculated phase diagram and the net low-T 
tunneling gap, defined as half the peak-peak separation.  The inset shows that 
in the mixed CDW-superconducting state a single gap evolves in the calculated 
tunneling density of states $\rho (E)$ (except for phonon structure).
The tunneling peaks coincide with the split electronic energy 
dispersion near $(\pi ,0)$ and $(0,\pi )$ of the Brillouin zone. 
These calculations demonstrate that a normal state pseudogap will show up in the
tunneling spectra.  This normal state gap need not be associated with a CDW, but
could be a flux phase\cite{Affl}, spin density wave\cite{KaSch}, or 
some more exotic phase. 
\par
All three papers\cite{tu1,tu2,tu3} report a prominent dip in the tunneling 
spectrum near twice the gap energy, suggested to be associated with an energy 
threshold for quasiparticle decay.  Such a decay channel should be mainly
sensitive to the tunneling density of states, regardless of the nature of the
pseudogap.
\par
Renner, et al.\cite{tu1} state that `the pseudogap is centered at the Fermi
level in both under- and overdoped samples.  It is therefore unlikely that the
pseudogap results from a band structure effect.'  This statement assumes that
doping is accomplished by a rigid band filling. However, it has been repeatedly 
observed that {\it strong correlations pin the Fermi level to a Van Hove
singularity (VHS) over an extended doping range}\cite{Surv}.  In
the calculations of Fig. \ref{fig:1}, the Fermi level was assumed pinned to the
VHS for doping up to $x$=0.125, with a fixed VHS and rigid band filling at
higher doping, to simulate these effects.  

We would like to thank NATO for enabling A.M. Gabovich to visit us and discuss
his work. Publication 733 of the Barnett Institute.

{\bf $*:$} On leave of absence from Inst. of Atomic Physics, Bucharest, 
Romania

\end{document}